# Radiomics-enhanced Deep Multi-task Learning for Outcome Prediction in Head and Neck Cancer

Mingyuan Meng[1], Lei Bi[1], Dagan Feng[1,2], and Jinman Kim[1]

[1] School of Computer Science, The University of Sydney, Sydney, Australia.
[2] Med-X Research Institute, Shanghai Jiao Tong University, Shanghai, China.
`lei.bi@sydney.edu.au`

**Abstract.** Outcome prediction is crucial for head and neck cancer patients as it can provide prognostic information for early treatment planning. Radiomics methods have been widely used for outcome prediction from medical images. However, these methods are limited by their reliance on intractable manual segmentation of tumor regions. Recently, deep learning methods have been proposed to perform end-to-end outcome prediction so as to remove the reliance on manual segmentation. Unfortunately, without segmentation masks, these methods will take the whole image as input, such that makes them difficult to focus on tumor regions and potentially unable to fully leverage the prognostic information within the tumor regions. In this study, we propose a radiomics-enhanced deep multi-task framework for outcome prediction from PET/CT images, in the context of HEad and neCK TumOR segmentation and outcome prediction challenge (HECKTOR 2022). In our framework, our novelty is to incorporate radiomics as an enhancement to our recently proposed Deep Multi-task Survival model (DeepMTS). The DeepMTS jointly learns to predict the survival risk scores of patients and the segmentation masks of tumor regions. Radiomics features are extracted from the predicted tumor regions and combined with the predicted survival risk scores for final outcome prediction, through which the prognostic information in tumor regions can be further leveraged. Our method achieved a C-index of 0.681 on the testing set, placing the 2[nd] on the leaderboard with only 0.00068 lower in C-index than the 1[st] place.

## 1 Introduction

Head and Neck (H&N) cancers are among the most common cancers worldwide (5[th] by incidence) [1]. Outcome prediction is a crucial task for H&N cancer patients in clinical practice, as it provides prognostic information for clinicians to guide treatment planning at an early stage so as to improve the survival outcomes of patients. However, outcome prediction is a challenging task as survival outcomes are intrinsically driven by many influential factors, such as clinical demographics, treatment regimens, and disease physiology [9]. In addition, outcome prediction has to take into account incomplete survival data. Generally, survival data includes many right-censored samples, for which the exact time of events occurring (e.g., disease progression or recurrence) is unclear. For example, patients may be lost to follow-up, or the events are not observed due to limited follow-up time. In these cases, it is only known that the events did not occur in a certain period of time and the events might occur later. Outcome prediction models, to make the maximum use of existing information, need to build from both complete (uncensored) and incomplete (censored) samples.

The HEad and neCK TumOR segmentation and outcome prediction challenge (HECKTOR 2022) invites the research community to develop algorithms for tumor segmentation and outcome prediction in H&N cancers using pretreatment PET/CT [2, 3]. There are two tasks in this challenge: Task 1 - segmentation of primary tumors and lymph nodes in PET/CT images, and Task 2 - prediction of patient outcomes from PET/CT images and available clinical data (e.g., gender, age, etc.). Participating algorithms should perform fully automatic inference from entire PET/CT images without requiring bounding boxes. In this study, we focus on Task 2 and intend to improve outcome prediction. As our method also produces segmentation masks of tumor regions (primary tumors and lymph nodes) as an intermediate output, we also participated in Task 1 and report the related experimental results in this paper.

Radiomics, referring to the extraction and analysis of handcrafted quantitative features, is widely adopted in clinical practice for outcome prediction [4-6]. However, radiomics methods require manual segmentation of tumor regions for every patient, which is intractable and error-prone. To address this limitation, deep learning methods based on deep neural networks have been proposed, which perform end-to-end outcome prediction without requiring manual segmentation [7-9]. However, without segmentation masks, these methods take the whole image as input, which makes them difficult to focus on tumor regions and potentially unable to fully leverage the prognostic information in the tumor regions [12]. Recently, deep multi-task learning was introduced for joint outcome prediction and tumor segmentation [10-12]. Through jointly learning with tumor segmentation, deep multi-task models are implicitly guided to extract features related to tumor regions, which relieves the above-mentioned limitation of focusing on tumor regions. In addition, as mentioned above, bounding boxes are not given in this challenge. Under this circumstance, a deep multi-task model that performs both outcome prediction and tumor segmentation could be an optimal solution as it removes the requirement for additional tumor detection.

In our previous study, we proposed a Deep Multi-task Survival model (DeepMTS) for joint outcome prediction and tumor segmentation, and it has shown promising performance in H&N cancers [11] and Nasopharyngeal Carcinoma (NPC) [12]. However, although the DeepMTS has been shown to have the ability to focus on tumor regions by deep multi-task learning [12], we found it still cannot fully leverage the prognostic information in tumor regions. In this study, we extend the DeepMTS to a radiomics-enhanced deep multi-task framework, where radiomics features extracted from DeepMTS-segmented tumor regions were incorporated as an enhancement. We demonstrated that this extension improved DeepMTS by a large margin.

## 2 Materials and Methods

### 2.1 Patients

The organizers of HECKTOR challenge provided a training set of 524 patients acquired from 7 centers, including CHUM (n = 56), CHUP (n = 72), CHUS (n = 72), CHUV (n = 53), MDA (n = 198), HGJ (n = 55) and HMR (n = 18). A testing set of 359 patients was provided for evaluation, in which 200 patients are from a center present in the training set (MDA) while the other 159 patients are from two centers absent from the training set (USZ and CHB). Note that the ground-truth labels of the testing set were not released to the public and we did not use any external dataset in this challenge. All patients were histologically confirmed with oropharyngeal H&N cancer and received radiotherapy and/or chemotherapy. Each patient underwent FDG-PET/CT before treatment and was recorded with clinical indicators including age, gender, weight, treatment regimens, Human Papilloma Virus (HPV) status, performance status, and tobacco/alcohol consumption. For Task 1 (tumor segmentation), primary Gross Tumor Volume (GTVt) and nodal Gross Tumor Volumes (GTVn) were annotated by experts and were regarded as ground-truth labels. For Task 2 (outcome prediction), Recurrence-Free Survival (RFS), including time-to-event in days and event status, was provided as ground-truth labels. The provided event status is a binary indicator, where 1 indicates patients with observed disease recurrence and 0 indicates censored patients. Since some patients did not exhibit complete responses to the treatment, only 489 and 339 patients in the training and testing sets have RFS labels. Therefore, our method was trained and validated using 5-fold cross-validation with 489 patients in the training set, and then was tested with 359 and 339 patients in the testing set for Task 1 and Task 2.

### 2.2 Radiomics-enhanced Deep Multi-task Framework

We propose a radiomics-enhanced deep multi-task framework that performs outcome prediction with PET/CT images and available clinical indicators. As shown in Fig. 1, the proposed framework adopts our recently proposed DeepMTS [12] as the backbone and then incorporates an automatic radiomics module as an enhancement. Specifically, DeepMTS is a deep multi-task model that can jointly predict the survival risk scores of patients and the segmentation masks of tumor regions (detailed in Section 2.3). With the segmentation masks derived from the DeepMTS, a radiomics module is used to extract discriminative features from the predicted tumor regions of PET/CT images (detailed in Section 2.4). We refer to this radiomics module as automatic radiomics, which differentiates it from traditional radiomics that relies on manual segmentation. Finally, the predicted survival risk scores, the discriminative radiomics features, and the clinical indicators are combined to build a Cox Proportional Hazard (CoxPH) model [13] for final outcome prediction.

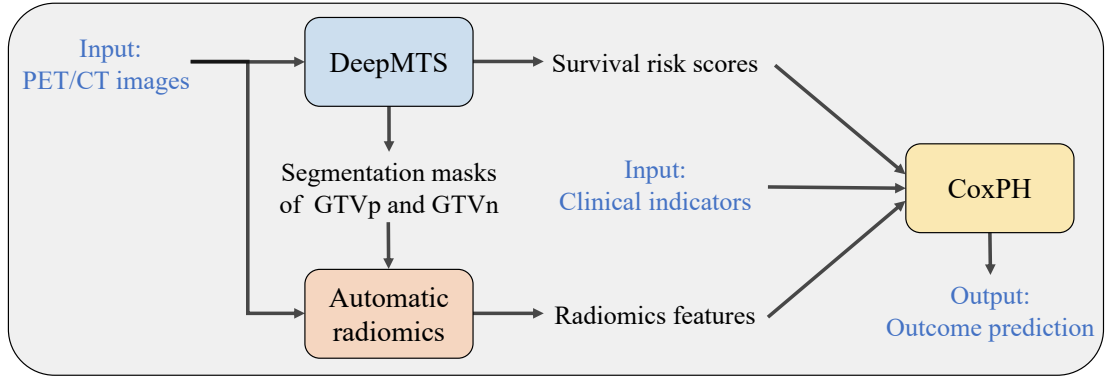

**Fig. 1.** The workflow of our radiomics-enhanced deep multi-task framework.

### 2.3 Deep Multi-task Survival Model (DeepMTS)

Fig. 2 shows the overall architecture of the DeepMTS [12], which consists of a U-net [14] based segmentation backbone, followed by a DenseNet [15] based cascaded survival network (CSN). The segmentation backbone is hard-shared by tumor segmentation and outcome prediction tasks, which implicitly guides the model to extract features related to tumor regions. The outputs of the segmentation backbone are fed into the CSN as a supplementary input, which provides the CSN with global tumor information (e.g., tumor size, shape, and locations). Deep features are derived from both the segmentation backbone and the CSN, and then are used to predict survival risk scores through several fully-connected layers.

The DeepMTS is trained in an end-to-end manner to minimize a combined loss $L$ of a segmentation loss $L_{seg}$ and a survival prediction loss $L_{surv}$:

$$L = L_{seg} + L_{surv} \qquad (1)$$

The $L_{seg}$ is the Dice loss [16] between the predicted probability maps and the ground-truth labels, while the Cox negative logarithm partial likelihood loss [17] is used as the $L_{surv}$ to handle right-censored survival data:

$$L_{surv} = -\frac{1}{N_{E=1}} \sum_{i:E_i=1} (h_i - \log \sum_{j \in \mathcal{H}(T_i)} e^{h_j}) \qquad (2)$$

where $h$ is the predicted risk scores, $E$ is the event status, $T$ is the time of RFS (for $E$ =1) or the time of patient censored (for $E$ =0), $N_{E=1}$ is the number of patients with disease recurrence, and $\mathcal{H}(T_i)$ is a set of patients whose $T$ is no less than $T_i$.

DeepMTS has two advantages when compared to employing two separate segmentation and outcome prediction models: (i) The DeepMTS can produce more accurate survival risk scores than single-task survival models [12]; (ii) Through jointly learning with outcome prediction, the segmentation target might become closer to the regions providing prognostic information [21], which will potentially improve the performance of downstream automatic radiomics. Further discussion and related ablation analysis can be found in our previous work [12].

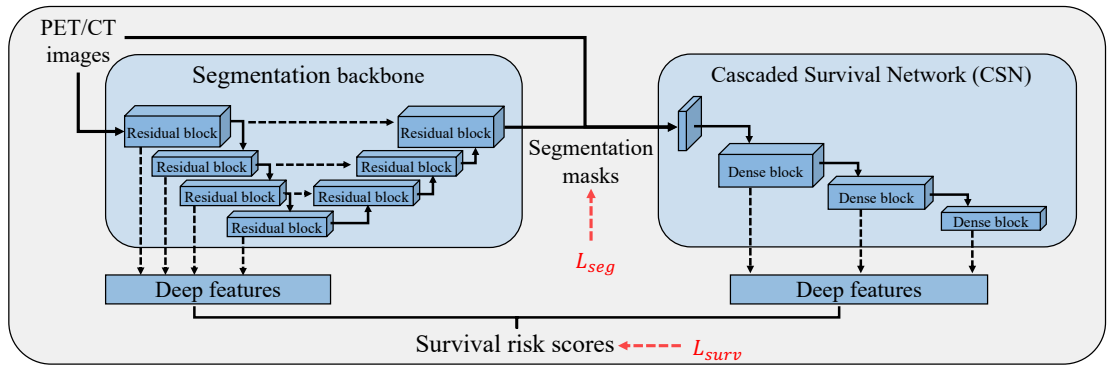

**Fig. 2.** The overall architecture of DeepMTS. More details can be found in [12].

### 2.4 Automatic Radiomics

For automatic radiomics, handcrafted features are extracted from the predicted tumor regions of PET/CT images by Pyradiomics [18]. The predicted GTVp and GTVn masks are merged into a single mask (1 for GTVp and GTVn and otherwise 0) for feature extraction. The extracted handcrafted features include the features from First Order Statistics (FOS), Neighboring Grey Tone Difference Matrix (NGTDM), Grey-Level Run Length Matrix (GLRLM), Grey-Level Size Zone Matrix (GLSZM), Grey-Level Cooccurrence Matrix (GLCM), and 3D shape-based features. In addition to the original PET/CT, eight wavelet decompositions of PET/CT are also used, resulting in a total of 1689 radiomics features. All radiomics features are standardized using Z-score normalization and then are reduced to seven discriminative radiomics features through Least Absolute Shrinkage and Selection Operator (LASSO) regression.

### 2.5 Image Preprocessing

We resampled images into isotropic voxels of unit dimension to ensure comparability, where 1 voxel corresponds to 1 mm$^3$. Trilinear and nearest neighbor interpolations were used for PET/CT images and segmentation masks, respectively. Based on the provided segmentation ground-truth labels, each image in the training set was cropped to 160×160×160 voxels with the tumor center aligning with the image center. However, as no bounding box was provided, each image in the testing set was cropped to 240×240×240 voxels to ensure the entire head and neck region can be included. For intensity normalization, PET images were standardized using Z-score normalization, while CT images were clipped to [−1024, 1024] and then mapped to [−1, 1].

### 2.6 Training and Inference

We first implemented the DeepMTS using Keras with a Tensorflow backend on two 12GB GeForce GTX Titan X GPUs. Specifically, we used an Adam optimizer with a batch size of 8 to train the DeepMTS for a total of 10,000 iterations. The learning rate was set as 1e−4 initially and then was reset to 5e−5 and 1e−5 at the 2,500[th] and 5,000[th] training iteration. Data augmentation was applied to the PET/CT images in real-time during training to minimize overfitting. The used data augmentation techniques include random affine transformations and random cropping to 112×112×112 voxels. Moreover, we sampled an equal number of censored and uncensored samples during the data augmentation process to minimize the problem introduced by the imbalanced censoring rate (79% in the training set). After the DeepMTS was trained, automatic radiomics was performed based on the segmentation outputs of DeepMTS. Finally, a CoxPH model was built based on the outputs of DeepMTS and automatic radiomics. During inference, the DeepMTS was used to locate the tumor region through a slide-window segmentation process. Then, the image patch containing the whole tumor region was fed into the proposed framework for outcome prediction.

### 2.7 Ensemble

Our results on the testing set were obtained using an ensemble of five models built with 5-fold cross-validation. For outcome prediction (Task 2), the testing results of the five models were first standardized by Z-score normalization and then averaged together to obtain the final testing results. For tumor segmentation (Task 1), we tried two ensemble approaches: (i) The testing results of the five models were first averaged together and then thresholded at 0.5 to obtain the final testing results, and (ii) The testing results of the five models were first thresholded at 0.5 and the final testing results were obtained through majority voting.

### 2.8 Evaluation Metrics

Tumor segmentation (Task 1) is evaluated using aggregated Dice Similarity Coefficient (DSC) [19]. The DSC metric measures the similarity between the predicted and ground-truth segmentation masks, which is computed separately for GTVp and GTVn. Outcome prediction (Task 2) is evaluated using concordance index (C-index) [20]. The C-index metric measures the consistency between the predicted survival scores and the ground-truth survival outcomes.

## 3 Results and Discussion

### 3.1 Outcome Prediction

Our experimental results of outcome prediction are reported in Table 1. For 5-fold cross-validation, the average results in five folds are reported along with the range in parentheses. We first evaluated the performance of using clinical indicators, automatic radiomics, and DeepMTS, separately. Then, we combined them in our radiomics-enhanced deep multi-task framework. The results of traditional radiomics (based on ground-truth manual segmentation) are also reported in Table 1 for comparison. We found that the DeepMTS achieved higher validation C-index than clinical indicators and radiomics methods, which is consistent with the results reported in our previous study [12]. Nevertheless, the DeepMTS still can be further improved by embedding clinical indicators and automatic radiomics. This demonstrates that, although the DeepMTS has been shown to have the ability to focus on tumor regions through deep multi-task learning [12], it still cannot fully leverage the prognostic information in the tumor regions and the automatic radiomics could be an effective enhancement. In addition, we found that automatic radiomics achieved higher validation C-index than traditional radiomics, which suggests that the provided ground-truth tumor regions might not be the optimal regions for prognosis. As we mentioned in Section 2.3, the regions predicted by DeepMTS may provide more prognostic information due to joint learning with outcome prediction.

During the challenge, only a maximum of 3 testing submissions were allowed. We, therefore, submitted the methods with the top-3 validation C-index. The highest testing C-index was achieved by combining DeepMTS, clinical indicators, and automatic radiomics, which made us place the 2$^{nd}$ on the leaderboard with a negligible difference from the 1$^{st}$ place (0.00068 lower in C-index: 0.68084 vs 0.68152).

Table 1. The C-index results on the 5-fold cross-validation and testing set.

| Method | 5-fold cross-validation | Testing set |
|---|---|---|
| Clinical indicators | 0.662 (0.613-0.689) | / |
| Traditional radiomics | 0.681 (0.625-0.732) | / |
| Automatic radiomics | 0.688 (0.644-0.735) | / |
| DeepMTS | 0.705 (0.663-0.740) | 0.644 |
| DeepMTS + Clinical indicators | 0.731 (0.702-0.773) | 0.649 |
| DeepMTS + Clinical indicators + Automatic radiomics | 0.768 (0.723-0.786) | 0.681 |

### 3.2 Tumor Segmentation

We also submitted the segmentation masks predicted by DeepMTS for Task 1. The testing results of using two different ensemble approaches (averaging and majority voting) are shown in Table 2. We found that averaging the outputs of five models (built with 5-fold cross-validation) achieved higher testing DSC than majority voting. We also found that the DeepMTS achieved higher testing DSC on GTVp than GTVn. This is because the DeepMTS was developed for primary tumor segmentation and we did not specially optimize it for lymph node segmentation. It should be noted that we focus on outcome prediction and the segmentation performance did not degrade our framework's performance in the outcome prediction. Instead, the predicted segmentation masks potentially improved the performance of radiomics (Table 1).

Table 2. The DSC results on the testing set.

| Ensemble approach | GTVp | GTVn | Mean |
|---|---|---|---|
| Averaging | 0.761 | 0.659 | 0.710 |
| Majority voting | 0.760 | 0.658 | 0.709 |

## 4 Conclusion and Limitations

In this study, we have outlined a radiomics-enhanced deep multi-task framework for outcome prediction in the context of HECKTOR 2022. With our recently proposed DeepMTS as the backbone, we incorporate automatic radiomics as an enhancement to further leverage the prognostic information available in the tumor regions. Our deep multi-task framework achieved a competitive result in the testing set of HECKTOR 2022, which made us place the 2$^{nd}$ in the leaderboard and only have a negligible difference (0.00068 lower in C-index) from the 1$^{st}$ place. Nevertheless, this study still has some limitations, and we suggest better performance potentially could be obtained by addressing the following:

Firstly, the balance between the losses of segmentation task and outcome prediction task was not fully explored. For simplicity, the segmentation loss $L_{seg}$ and the survival prediction loss $L_{surv}$ were directly summed without using any weighting parameters (equal weight was assigned to $L_{surv}$ and $L_{seg}$), which might lead to sub-optimal performance. Secondly, the DeepMTS relies on an early fusion strategy where CT and PET images are concatenated as a 2-channel input. Other fusion strategies, such as intermediate or late fusion, could be further explored and this might enable better performance. Finally, due to limited time and computing resources, we performed 5-fold cross-validation to build the final model ensemble. However, leave-one-center-out validation might be a better alternative as this enforces models to learn prognostic information across different centers, which is helpful to build a robust ensemble.